# (Science) Teacher Education in the Knowledge Society: Good Practice at the University of Geneva

Andreas Müller[a], University of Geneva, 6/2018

The present text summarizes several factors considered as essential for teacher education according to the state of current research and best practice, with a focus on science teacher education, and as realized at the University of Geneva.

## 1    Teacher education *at universities:* Professional Schools of Education and other structural elements

In this section, it is argued that teacher education *at universities* has a natural, if not necessary place in an educational system well prepared for the future, and what institutional and structural elements may support such a role – in general, and in particular for Science teacher education.

The "University Institute of Teacher Education (IUTE[b])" of the University of Geneva is a "Professional School of Education (PSE)". At the international level, such a structure is considered as one of the best approaches for education (initial and continuing) of teachers, in line with internationally recognized models like the PSE at Stanford [1], [2]. In short, the main objective of a PSE is to offer a teacher education structure that integrates the fundamental elements of *research – (academic) teaching – (school) practice*. For this purpose, a first basic feature of a PSE is an academic education both oriented towards the practice of the teaching, and based on research both in education and in the taught disciplines. In knowledge societies, the education of teachers, as best guarantors of general and scientific literacy of future citizens, must have the best available foundations in the scientific knowledge, and also in the attitudes and approaches of how to develop and confirm it – in short, in the best available foundations in research. Universities are the best institutions to ensure such a strong foundation in research (taught disciplines, educational sciences). The second guiding idea is to achieve a unity of teaching and research [3], [4], which is a well-established standard for the vast majority of academic disciplines, but too often neglected in teacher education. Specifically, this requires to establish discipline-based education research (such as science education research, [5]) at the faculty level (as achieved at the IUTE of Geneva with 8 professors for various disciplinary education research, with 2 for science education). Finally, the third basic feature is to bring together the disciplines necessary for teacher education (educational sciences, discipline-based education) in a common interfaculty structure, with dual affiliation of the chairs to their "primary" faculty. This is a good foundation to ensure coherence and appropriate coordination between the components of teacher education and a vivid academic connection with their respective disciplines[2)]. Together, these features do not automatically guarantee success, as teacher education is a multifactorial and sometimes quite controversial task with many stakeholders. But it is, to the best of present knowledge, one of the solutions with the highest chances and potential on the structural level [2].

At an institutional level, the PSE in Geneva is part of the second largest university in Switzerland, which enjoys an international reputation, in particular for science. As such, the IUTE is the only institution of this kind in Switzerland, and at the European level it belongs to a group of a PSE at university with a high standard of research, such as the "Technical University of Munich" [6].

Three further specificities in Geneva are important for an effective teacher education. *First*, the integration of initial and continuous education in the same institution, which is the only way to assure the integration of research – education – practice as essential strength of a PSE also for in-service teachers. This is crucial for two reasons: (i) The initial in-service years are known to be decisive for the development of the professional attitude of teachers. (ii) At any time, the vast majority of teachers are obviously in-service teachers, who have to teach live-long learning to their pupils, and in order to do so, have to practice it themselves. Thus, both for the critical initial phase, and for live-long learning, continuous professional education is best placed in a PSE, if it is to be based on an effective synthesis of research, education and practice. *Second*, an understanding of school education and thus of teacher education as a coherent whole, from primary school to secondary II. An important characteristic of this is that the teaching qualification is awarded simultaneously for secondary level I and II. This strongly fosters a good understanding of what is done on both levels, and of the transition between them Actually, many teachers are employed on both levels. (The possibility of this is also a considerable advantage with respect to the labour market.) Another important aspect is the link with primary education. It is well known, that the basis for science interest and knowledge is laid in primary school, where however most teachers do not have a solid science background. The cooperation between teacher education for the primary and secondary levels in the same institution can help in that respect. For science there is a contribution for primary teacher education on a regular basis, and other forms of cooperation within science education projects. *Third*,

---

the good integration of science education as academic discipline into the science faculty, and the strong commitment of the latter for educational matters (important e.g. as valuable resource for initial, continuous and lifelong learning of science teachers, as mentioned above).

In this respect, a specific strength for science education and science teacher education in Geneva is the existence of a very broad and high quality spectrum of out-of-school learning offers established by the faculty of science, most notably the Scienscope [7]. Its objective is to foster a positive attitude and understanding of science among the population in general, and young people in particular. It offers entertaining and fascinating workshops on more than 50 topics for more than 10000 participants a year. Disciplines included are biology, chemistry, computer science, mathematics, and physics, and soon the earth sciences. Moreover, Bioutils [8], providing kits and expert support on topics of modern biology for the classroom, and the Stellarium Gornergrat [9], a remote-accessible observatory for educational purposes have to be mentioned here. Moreover, there is a specific platform for keeping alive the contact of science teachers with "their" science, called experiment@l [10]ü, and supported by the faculty of science and the DIP. All these offers are also opportunities where teacher education (initial and continuous) using a unique approach to experience the sciences takes place, and where a vivid exchange with in-service teachers helps to keep alive their own contact with and interest for their discipline.

Science teacher education along these lines runs well in Geneva. The accreditation report for the IUTE acknowledges the realization of the guiding ideas mentioned above, in particular research-based teacher education and the good coordination between theory and practice. Evaluation of initial science teacher education courses are good to very good. This is due, among other factors, to the evidence-based approach in education, as described in sect. 2. Another indicator is a very low number of science teacher students not successfully finishing theirs studies. For continuous professional education, there is a rich offer by collegues from science and science education, with good attendance by teachers. Finally, there are several cases, where in-service teachers who graduated at the IUFE later on seek contact with the science education group and get involved in the teacher education program for the younger collegues, and propose and participate in research and development projects in science education.

The Faculty of Science of the University of Geneva, with the structures, approaches and commitment as described above, and with about 30% of its graduates working in the teaching profession later on (thus being its largest job sector, [11]), is well-placed to be a major acting force for science teacher education, so important for our society. However, the claim is not that these specificities are realized in an ideal way in Geneva, but that they work as important elements of good practice for a successful science teacher education, potentially useful also in other circumstances.

## 2 Evidence-Based Education

### 2.1 Background

> *The remarkable feature of the evidence is that the greatest effects on student learning occur when teachers become learners of their own teaching.*
> Hattie, J. (2012). Visible learning for teachers [12]

Evidence-based practice is the approach to base decisions on the best available evidence, in the sense of the best possible – in particular systematic! – use of existing knowledge and research. The earliest example in this respect was evidence-based medicine (EBM), i.e. "[t]he conscientious, explicit and judicious use of current best evidence in making decisions about the care of individual patients", as one of its pioneers put it [13]. The common ground with evidence based education is that a precious value and objective – good health, or good education – has to be realized with limited means. This has led to a very strong current of research and practical implementation in Evidence-Based (Science) Education (EB(S)E) in the last two decades ([14], [15], [16]). The work of Hattie ([12], [17]) based on more than 800 meta-analyses (comprising more than 80 000 000 individuals) is considered to be groundbreaking in this area. Other highly influential sources are the "Best Evidence Encyclopedia" of the John Hopkins University [18], and work done at the Stanford PSE, mentioned above [19].

By way of example, results of general importance here are the considerable effect sizes for achievement by providing formative evaluation (Cohen $d = 0.90$[2)]; [17]) and by cooperative vs. individualistic learning ($d = 0.64$; [20]). Other important research is about affective and motivational aspects, eg. about interest and self-efficacy believes [21], or curiosity [22].

A further informative example is about homework, a topic strongly debated among parents, teachers, and researchers: overall effects are between small and medium size[2)] ($d$ from 0.36 to 0.65), but a much better image is obtained when taking account of feedback as co-factor, which when given, or not, leads to a large contrast between strong and weak effects ($d = 0.83$ vs. $d = 0.28$; [26]). This is an illustrative example of three general statements: (*i*) EBE and meta-analyses can only be as good as important co-factors have been taken into account; simplistic recipes are not a goal one is looking for. (*ii*) The co-factor in question here, feedback, is consistent with the strong positive effects for formative evaluation (a kind of systematic feedback) mentioned above. As a matter of fact, development of improved methodology allows for research synthesis aiming at this kind of more detailed analysis, and going increasingly beyond classical meta-analysis ([27], [28], [29]). (*iii*) Last but not least, one sees how EBE can help to prevent from fruitless debates (such as about homework), hitherto based all too often on mere conviction and ideology.

In the sense of Hattie´s quote above, EBE is the approach to provide a systematic and conscientious base of best evidence for teachers to "become learners of their own teaching". As a word of caution, however, it has to be mentioned that there is a thorough debate about EBE (as always in educational matters), and it is certainly not its claim to guarantee all by itself a solution to all problems in science education and science teacher education, as evidenced by the following statement by Millar et al. [30]: „We need to work towards a situation in which research evidence is routinely an explicit input to teachers' decision making, but where it is also accepted that this must be weighed judiciously alongside other kinds of knowledge to reach a decision that can be rationally defended." It is such a well-balanced point of view we subscribe to in the approach described here.

In science education, and science teacher education, in particular, the evidence-based approach has seen a very intensive development. It is among the strongest ones for educational research on school subject matters, (together with mathematics education), and the idea has found strong support among many scientists interested in effective teaching and learning of their disciplines, also on highest reputational level [31] - [33]. In particular, the physics Nobel price winner C. Wieman has devoted his work in the last decade to science education research (since 2013 at the PSE in Stanford; [34] - [36]). Several illustrative findings for three aspects might serve as example, both on the cognitive and the affective level, see following sections.

*2.2 Learning effectiveness*

> [R]eform in science education should be founded on "scientific teaching",
> in which teaching is approached with the same rigor as science at its best
> Handelsman, J., et al. (2004), Science, 304, 521 [31]

For improving learning effectiveness, a first illustrative example are collaborative learning strategies. When applied to science learning, they yield strong positive effects ($d = 0.95$; [37]), consistent with, and even more pronounced than the effects found across all disciplines (see above). Another example highly relevant for science education is inquiry-based learning (IBL), which finds strong support on the political level [38]. However, there is little empirical support for the effectiveness of IBL ($d = 0.31$; [17]). Looking more closely, a meta-analysis on IBL in science by Furtak et al. [39] has shown that teacher guidance is an decisive moderator: effect sizes are more than twice as large with guidance than without ($d = 0.65$ and $d = 0.25$, respectively). Unguided inquiry, as IBL is often understood, is thus *not* an effective approach for science learning. As in the case of homework, a differentiated, quantitatively based stance has to replace a simplistic, and sometimes ideological one. Research syntheses and meta-analyses for the "big picture" of EBSE are available ([37], [40], [41]) and provide a sound basis for science education.)
1) An interesting example for results on a more fine-grained level is the improvement of science literacy and understanding by inclusion of astronomy topics, in the classroom or in planetariums, from primary level pupils to teacher students. Results show positive effects for general science literacy (eg. the origin of seasons or the motion of Earth and Moon) or more specific competencies (eg. spatial abilities), with many effect sizes $\geq 0.8$. .

Tab. 1 in note 3) below gives an overview of some selected results, as an example of how EBSE can inform about quite specific questions. We will come back to astronomy in the next paragraph.

*2.3 Raising motivation and interest*

> This has been the most exciting thing I have done on this degree so far
> Primary teacher student using a remote access telescope
> in an introductory astronomy course [42]

A very important example here are hands-on-activities, which are considered by many practitioners as a strongly motivating element for learners. For science learning, a large-sample study ($N \approx 8000$, age group 11-16 a) has indeed shown, that they are the type of learning activities meeting the highest degree of interest (high/very interest: $\geq 50\%$ of learners ([41]). A recent study ([43]) confirms hands-on activities as strong predictors of science interest at school. Moreover, another large-sample study conducted by the [44] cooperation has shown a large impact of hands-on learning also on science achievement ($d = 0.91$). It is clear, that science teachers have to be aware of this and other effective means of fostering interest and learning among pupils, and also about their relative strengths. Reviews and theoretical synthesis of this and other important affective factors exist and provide a good basis for science teacher education [45].

While the impact of hands-on approaches on science interest probably does not come as a surprise (even though knowing "how much" in the spirit if EBSE is certainly superior to merely knowing "that"), there are more recent developments where evidence about their impact is useful. One important case of this are out-of-school learning offers like the Scienscope, Bioutils and the Stellarium Gornergrat mentioned above. A considerable body of research indeed supports their great motivational potential (research journals: [46] – [49]; highly influential sources, such as the National Research Council [50], and reference works: [51]). As Rennie [52] puts it, "an enjoyable and successful visit experience is an important outcome because it can predispose the learner to engage in further cognitive learning. Motivation and willingness to engage in further instruction are most likely to be the important affective outcomes of a visit." For instance, a series of studies on more than 10 sites has shown quite positive results for enjoyment/general appreciation of various out-of school science learning offers (70-90% of maximal value), consistently across several countries (France, Germany, Switzerland), age groups (primary to secondary level II), various settings (single and multiple visits, degree of guidance) and disciplines (biology, chemistry, physics), and across more than a decade [53].

EBSE provides also for more detailed, partially unexpected results, such as that among all areas of science, astronomy topics are among the most interesting ones for young people, much more interesting than many conventional school topics (ROSE – Relevance in Science Education; [54], [55]). Similar results were found in several different countries, within and outside the scope of the ROSE study (eg. within ROSE: Austria, England, Finland, Germany, Ireland, Israel, Sweden; see [54]). In an independent investigation, Osborne & Collins [56] state the central message as follows: *Aspects of the subject that 'amazed' or 'fascinated' were limited to those topics that had personal relevance, either to their everyday lives or those that dealt with existential questions of identity such as astronomy and cosmology*. Note that these findings are particularly important in view of the fact that the investigated pupils were of an age where decisions about their future school and professional development are about to be taken. Moreover, the strong motivational potential of astronomy as science topic applies also for teacher education, as evidenced by the statement at the beginning of this section.

In view of the notable positive effects both for both interest and learning (see above) there are strong reasons to include astronomy in science teaching at school and a fortiori in science teacher education. In the science teacher education program in Geneva, this is taken account of in several compulsory teaching units of initial education and also by a series of continuous professional education offers. One question to be solved here is that astronomy is rarely taught as a school discipline as such, and one has to find ways for an embedding in existing curricula (see note 0. It is here that an offer like the Stellarium Gornergrat (see above) gains its full weight. First, it offers first hand access to astronomical images and data of unprecedented quality for educational purposes. Second, learning activities developed and tested by participating teachers ensure practical feasibility and an adequate embedding in the local curricula.

### 2.4 Measurement

> *Figuring out what to measure, and how well to measure it, is critical in all fields. [...]*
> *Although the specifics for how to do this are different between physics and education,*
> *the basic methods are much the same.*
> C. Wieman, in [35]

Finally, the very idea of EBE is obviously based on the possibility to reliably measure changes and outcomes of educational variables, for cognitive, motivational, and other variables. The development of instruments for educational diagnosis, measurement and assessment has received sustained attention over the past two decades. Among other reasons, there is an increasing awareness for the need of educational experimentation tested by reliable measurements, very much in the sense of EBE ([35], [57] - [59]). Teachers have to know, whether a given approach really increases understanding or motivation, and they have to be able to test this in their own classrooms. A field of particular interests are concept tests, probing for "intuitive conceptions" of "misconceptions", such as the ideas that the seasons are caused by different distances to the sun, that a moving object looses force, or that dolphins are fish. Such ideas are found to be widespread, and hard to change in many science fields, even among university students (see eg. [33], [60], [61]). Teachers have to be aware of these learning difficulties, of ways how to overcome them, and again to diagnose them and look for possible change in their own classrooms.

In sum, measurement instruments have an essential role in science education research, practice, and teacher education, and will certainly become even more important in the future.

# 3   Conclusion

Teacher education is a central part of an educational system: as studies on university level, it is the last stage of an educational trajectory, and at the same time is the basis and starting point for the educational quality for many generations of school students. If universities, as strongholds of culture and knowledge in a society, want to ensure the transmission of their values and knowledge, in order to have good students and to ensure the academic succession, but also for the development of society in general, they have to strongly engage themselves for teacher education. A promising structure for this is a professional school of education, but other approaches do exist. For these goals, a Professional School of Education appears to be a promising institutional structure, and Evidence-Based Education to provide the most reliable knowledge basis. As stated several times, it is not claimed that the suggestions made here, and their realization in Geneva are ideal ones, or the only approach[2], but they are proposed as an example of good practice, possibly useful also for the discussion in other settings.

**Notes/supplementary information**

2) The present contribution is written from the point of view of a PSE at a general, research-oriented university. It does not discuss (nor exclude) specialized teacher education universities, for which in fact promising formats similar to the PSE discussed here exist, in particular when in close cooperation with a general university.

3) The basic definition is Cohen $d = (M_T - M_C)/SD$, where $M_T$ and $M_C$ are the means (of some variable of interest) for the treatment and control group, respectively, and $SD$ is the either the pooled standard deviation or that of the control group [23]. In simple terms, $d$ thus measures the impact of an intervention is units of standard deviations of the sample under consideration. Usual effect-size levels (as established from comparison of a great many of studies in different areas) are small ($0.2 < d < 0.5$), medium ($0.5 \leq d < 0.8$) or large ($0.8 \leq d$) [23]. Many modifications and refinements of the concept of "effect size" haven been developed and are used in the literature, see eg. [24] for an overview.

4) Tab. 1 (as an example of how EBSE can inform about quite specific questions): Impact of inclusion of astronomy topics on science literacy and understand in various settings and for various target groups (PP, SIP, SIIP: primary, secondary level I and II pupils, PTS: primary teacher students); $d$: comparison to a control group (regular classroom), $d^*$: post-pre comparison

| $d$ | learning objectives | setting | source |
|---|---|---|---|
| | **basic astronomy, general science literacy** | | |
| 0.75* | seasons (and other topics) | planetarium; SIP | [62] |
| 0.9*/1.0* | Earth and Moon | planetarium; SIP/ SIIP | [63] |
| 1.1*/0.8* | basic astronomy, science literacy (different topics) | science centre; PP/SIP | [64] |
| 0.65*-2.2* | celestial motion (daily, lunar and annual) | planetarium, PP | [65] |
| | **focus on specific competencies** | | |
| 1.0 | cosmic distances | planetarium; PP&SIP | [66] |
| 1.34 | spatial abilities; PTS (preparing for and guiding planetarium sessions) | planetarium; PTS (s.left) | [67] |
| | **integrated in classroom science course** | | |
| $d_T^*$=1.0 | physical science through astronomy activities time, light & color, earth in motion | classroom; PP | [68] |

Embedding of astronomical tropics in existing curricula is possible in several ways: First, many science curricula contain explicitly basic astronomy concepts and knowledge (primary and secondary level I: day/night, lunar and seasonal cycles, eclipses, basic structure of the solar system, galaxies, the universe etc.; secondary level II: Gravitation, Kepler laws). Second, many topics of the regular curriculum for several disciplines (in particular physics, mathematics, geography) can be taught with astronomical examples and activities, benefitting from their large motivational potential, for example: basic circle geometry, concept of velocity: activity about the course of stars in the night sky; on a more advanced level trigonometry: activity about the height of the lunar mountains (+ links to history); trigonometry, concept of angular velocity: activity about Jupiter's big red spot. Moreover, there are compulsory courses like "introduction to the scientific method" in the curricula for secondary level I and II where important contents are linked almost naturally to astronomical examples and activities (orders of magnitude, data evaluation, modelling, etc.).

---

[c] websites as accessed on 1/2/2018.